\documentclass[aps,prl,superscriptaddress,twocolumn]{revtex4}
\usepackage{graphics}
\usepackage{amsmath}
\usepackage{graphicx}
\usepackage{amsfonts}
\usepackage{amssymb}

\begin{document}

\title{Eavesdropping of two-way coherent-state quantum cryptography\\
via Gaussian quantum cloning machines}
\author{Stefano Pirandola}
\affiliation{M.I.T. - Research Laboratory of Electronics, Cambridge MA 02139, USA}
\author{Stefano Mancini}
\affiliation{Dipartimento di Fisica, Universit\`{a} di Camerino, I-62032 Camerino, Italy}
\author{Seth Lloyd}
\affiliation{M.I.T. - Research Laboratory of Electronics, Cambridge MA 02139, USA}
\affiliation{M.I.T. - Department of Mechanical Engineering, Cambridge MA 02139, USA}
\author{Samuel L. Braunstein}
\date{\today }
\affiliation{Computer Science, University of York, York YO10 5DD, United Kingdom}

\begin{abstract}
We consider one of the quantum key distribution protocols recently
introduced in Ref.~[Pirandola et al., Nature Physics \textbf{4}, 726
(2008)]. This protocol consists in a two-way quantum communication between
Alice and Bob, where Alice encodes secret information via a random
phase-space displacement of a coherent state. In particular, we study its
security against a specific class of individual attacks which are based on
combinations of Gaussian quantum cloning machines.
\end{abstract}

\maketitle

\section{Introduction}

Recently \cite{Nat,SDiego}, we have shown how two-way quantum communication
can profitably be exploited to enhance the security of continuous variable
quantum key distribution \cite{Ralph1,Preskill,Grangio,Christ}. In
particular, we have investigated the security of two-way protocols in the
presence of collective Gaussian attacks which are modelled by combinations
of entangling cloners \cite{Grangio}. Even though this situation is the most
important one from the point view of the practical implementation, the
effect of other kind of Gaussian attacks (i.e., not referable to entangling
cloners) must also be analyzed. In this paper, we study the security of the
two-way coherent-state protocol of Ref.~\cite{Nat} against individual
attacks where an eavesdropper (Eve) combines two different Gaussian quantum
cloning machines (also called \textit{Gaussian cloners}). In particular, we
are able to show the robustness of the two-way protocol when the first
cloner is fixed to be symmetric in the output clones. This symmetry
condition enables us to derive the results quite easily but clearly
restricts our security analysis to a preliminary stage. For this reason, the
\textit{optimal} performance of Gaussian cloners against two-way quantum
cryptography is still unknown at the present stage.

\section{Additive Gaussian channels and Gaussian cloners}

Consider a stochastic variable $X$ with values $x\in \mathbb{R}$ distributed
according to a Gaussian probability
\begin{equation}
G_{\Sigma ^{2}}(x)=\frac{1}{\sqrt{2\pi \Sigma ^{2}}}\exp \left[ -\frac{x^{2}%
}{2\Sigma ^{2}}\right] \,,  \label{real_Gauss}
\end{equation}%
with variance $\Sigma ^{2}$. This variable is taken as input of a classical
channel that outputs another stochastic variable $Y$ with values $y\in
\mathbb{R}$. In particular, the classical channel is called \textit{additive
Gaussian channel} if, for every input $x$, the conditional output $y|x$ is
Gaussianly distributed around $x$ with some variance $\sigma ^{2}$ \cite%
{Note}. As a consequence, the output variable $Y$ is a Gaussian variable
with zero mean and variance $\Sigma ^{2}+\sigma ^{2}$. According to
Shannon's theory \cite{Shannon}, the classical correlations between the
input and output variables lead to a mutual information
\begin{equation}
I(X,Y)=\frac{1}{2}\log (1+\gamma )~,
\end{equation}%
where $\gamma \equiv \Sigma ^{2}/\sigma ^{2}$ is the signal to noise ratio
(SNR). This formula gives the maximal number of bits per Gaussian value that
can be sent through a Gaussian channel with a given SNR (on average and
asymptotically).

In quantum information theory, an example of additive Gaussian channel is
provided by the Gaussian quantum cloning machine (GQCM) \cite{GQCM}.
Consider a continuous variable (CV) system, like a \emph{bosonic mode},
which is described by a pair of conjugate quadratures $\hat{x}$ and $\hat{p}$%
, with $[\hat{x},\hat{p}]=i$, acting on a Hilbert space $\mathcal{H}$. Then,
consider a coherent state $|\varphi \rangle $ with amplitude $\varphi
=(x+ip)/\sqrt{2}$. A $1\rightarrow 2$\ GQCM is a completely-positive
trace-preserving linear map
\begin{equation}
M:|\varphi \rangle \langle \varphi |\rightarrow \rho _{12}\in \mathcal{D}(%
\mathcal{H}^{\otimes 2})~,
\end{equation}%
such that the single clone states, $\rho _{1}=\mathrm{tr}_{2}(\rho _{12})$
and $\rho _{2}=\mathrm{tr}_{1}(\rho _{12})$, are given by a Gaussian
phase-space modulation of the input state $|\varphi \rangle \langle \varphi
| $, i.e.,%
\begin{equation}
\rho _{k}=\int \,d\mu \,\Omega _{\sigma _{k}^{2}}(\mu )\hat{D}(\mu )|\varphi
\rangle \langle \varphi |\hat{D}^{\dag }(\mu )\,,\quad k=1,2,\,
\label{Gauss_clon}
\end{equation}%
where
\begin{equation}
\Omega _{\sigma _{k}^{2}}(\mu )\equiv \frac{1}{\pi \sigma _{k}^{2}}\exp %
\left[ -\frac{\left\vert \mu \right\vert ^{2}}{\sigma _{k}^{2}}\right] \text{
,}  \label{Gamma}
\end{equation}%
and
\begin{equation}
\hat{D}(\mu )=\exp (\mu \hat{a}^{\dagger }-\mu ^{\ast }\hat{a})~.
\end{equation}%
In Eq.~(\ref{Gamma}), the quantities $\sigma _{k}^{2}$ are the error
variances induced by the cloning process on both the $x$ and $p$ quadratures
of the $k$-th clone. Notice that here we consider a GQCM which clones
symmetrically in the quadratures (in general, one can have a Gaussian cloner
which is asymmetric both in the clones and the quadratures, with four
different noise variances $\sigma _{1,x}^{2},$ $\sigma _{1,p}^{2},$ $\sigma
_{2,x}^{2}$ and $\sigma _{2,p}^{2}$.) The previous variances do not depend
on the input state (\textit{universal} GQCM) and satisfy the relation%
\begin{equation}
\sigma _{1}^{2}\sigma _{2}^{2}\geq 1/4~,
\end{equation}%
imposed by the uncertainty principle. In particular, the previous GQCM is
said to be \emph{optimal} if $\sigma _{1}^{2}\sigma _{2}^{2}=1/4$. In terms
of Shannon's theory, each of the two real variables, $x$ and $p$, is subject
to an additive Gaussian channel with noise equal to $\sigma _{k}^{2}$ during
the cloning process from the input state to the output $k$-th clone.

\section{Two-way coherent-state protocol}

\begin{figure}[h]
\vspace{-0.6cm}
\par
\begin{center}
\includegraphics[width=0.50\textwidth]{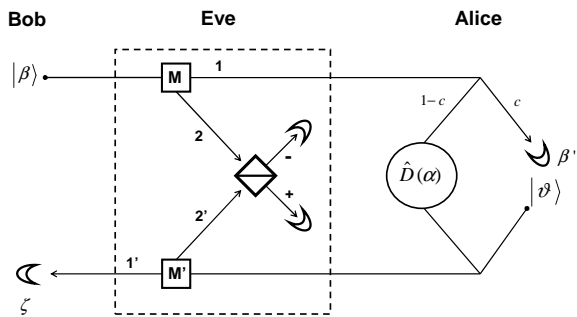}
\end{center}
\par
\vspace{-1.3cm}
\caption{Two-way coherent-state protocol in both the ON and OFF
configurations.}
\label{fig1}
\end{figure}

The protocol is sketched in Fig.\ref{fig1} and consists of two
configurations, \textit{ON} and \textit{OFF}, that can be selected by Alice
with probabilities $1-c$ and $c$ respectively.

Let Bob prepare a \emph{reference} coherent state $|\beta \rangle \langle
\beta |$, with amplitude $\beta $ randomly chosen in the complex plane
(e.g., according to a Gaussian distribution with a \emph{large} variance).
Such a state is sent to Alice on the forward use of the quantum channel. In
the ON\ configuration, Alice encodes a \emph{signal }on this reference state
via a phase-space displacement $\hat{D}(\alpha )$ whose amplitude $\alpha
\equiv (x_{A}+ip_{A})/\sqrt{2}$ is chosen\ in the $\mathbb{C}$-plane
according to a random Gaussian distribution $\Omega _{\Sigma ^{2}}(\alpha )$
with \emph{large} variance $\Sigma ^{2}$. Notice that

\begin{description}
\item[(i)] The \emph{signal }amplitude $\alpha $ symmetrically encodes two
signal quadratures, $x_{A}$ and $p_{A}$, i.e., two independent and real
random variables distributed according to Gaussian distributions $G_{\Sigma
^{2}}(x_{A})$ and $G_{\Sigma ^{2}}(p_{A})$.

\item[(ii)] The output state
\begin{equation}
\hat{D}(\alpha )|\beta \rangle \langle \beta |\hat{D}^{\dag }(\alpha
)=|\alpha +\beta \rangle \langle \alpha +\beta |~,
\end{equation}%
encodes the signal amplitude $\alpha $, masked by the reference amplitude $%
\beta $ chosen by Bob.
\end{description}

\noindent The state is finally sent back to Bob, who tries to guess the two
Alice's numbers $x_{A}$ and $p_{A}$ by a joint measurement of conjugate
observables \cite{Arthurs}. This is accomplished by a heterodyne detection
\cite{Shapiro} of the state, which will give an outcome $\zeta \approx
\alpha +\beta $. After the subtraction of the known value $\beta $, Bob
achieves an estimate $\alpha ^{\prime }$ of Alice's complex amplitude $%
\alpha $, i.e., $x_{A}^{\prime }\approx x_{A}$ and $p_{A}^{\prime }\approx
p_{A}$.

In the case of a noiseless channel between Alice and Bob, the only noise in
all the process is introduced by the heterodyne detection. This measurement
can be seen as a further Gaussian additive channel at Bob's site, which
gives a Gaussian noise equal to $1$ for each quadrature. Thus, according to
Shannon's formula, we have
\begin{equation}
I_{AB}=I(x_{A},x_{A}^{\prime })+I(p_{A},p_{A}^{\prime })=\log (1+\gamma
_{AB})~,
\end{equation}%
with $\gamma _{AB}=\Sigma ^{2}/1$.

Let us now consider a noisy channel adding Gaussian noise with variances $%
\sigma ^{2}$ (in the forward path) and $\sigma ^{^{\prime }2}$ (in the
backward path) for each quadrature. Then, the total noise of the channel is $%
\sigma _{ch}^{2}=\sigma ^{2}+\sigma ^{^{\prime }2}$ and the total noise
which Bob tests, after detection, is equal to $\sigma _{B}^{2}=\sigma
_{ch}^{2}+1$, giving a SNR $\gamma _{AB}=\Sigma ^{2}/\sigma _{B}^{2}$. In
the OFF configuration, Alice and Bob estimate the noise in the channel by
performing two heterodyne detections. After receiving the reference state,
Alice simply heterodynes it with outcome $\beta ^{\prime }$ and then
reconstructs a coherent state $\left\vert \vartheta \right\rangle $. This
state is sent to Bob, who gets the outcome $\zeta \approx \vartheta $ after
detection. In this way, Alice and Bob collect the pairs $\{\beta ,\beta
^{\prime }\}$ and $\{\vartheta ,\zeta \}$ from which they can estimate the
two noises $\sigma ^{2}$ and $\sigma ^{\prime 2}$ of the channel via public
communications. Notice that here we are using the ON configuration to encode
the key and the OFF configuration to check the noise of the channel. This
means that we are implicitly assuming that Eve's attack is disjoint between
the two paths of the quantum communication (i.e., Eve is using two distinct
one-mode GQCMs). More generally, in order to exclude joint attacks between
the two paths, the ON and OFF\ configurations must be used symmetrically for
encoding and checking \cite{Nat}.

\section{Eavesdropping via Gaussian cloners}

In the previous two-way quantum communication, the choices of the reference $%
\beta $ and the signal $\alpha $ are two independent processes. As a
consequence, Eve has to extract information on both the reference $\beta $
and the total displacement $\alpha +\beta $ in order to access Alice's
encoding $\alpha $ (this is true until the attack is disjoint). Let us
consider two different attacks, one on the forward use of the channel and
the other one in the backward use, by using two optimal GQCMs which we call $%
M$ and $M^{\prime }$, respectively (see Fig.~\ref{fig1}).

Since the reference $\beta $ and the signal $\alpha $ are chosen with large
variances, such machines must be universal, and since the information is
symmetrically encoded in the two quadratures, we consider equal cloning
noises in $x$ and $p$. For these reasons, Eve's GQCMs are exactly of the
kind specified by Eq.~(\ref{Gauss_clon}) with $\sigma _{1}^{2}\sigma
_{2}^{2}=1/4$. After cloning, Eve must extract the information about $\alpha
$ from her clones. She can directly heterodyne the clones. Alernatively, she
can send the clones to a beam-splitter (BS), with suitable reflection and
transmission coefficients $r$ and $t$, and then heterodynes the output ports.

In order to study the eavesdropping depicted in Fig.~\ref{fig1}, it is not
sufficient to consider the reduced states $\rho _{k}$ of the two single
clones at the output of $M$, but we have to compute explicitly the whole
bipartite state $\rho _{12}$\ of modes $1$ and $2$. In fact, mode $1$ is
sent to Alice (who displaces it) and then cloned by $M^{\prime }$ into the
output modes $1^{\prime }$ and $2^{\prime }$. The second mode $2^{\prime }$
then interferes with the previous mode $2$ on the beam-splitter. For this
reason, we have to keep all the correlations between the various modes till
the interference process. One can prove that the bipartite state $\rho _{12}$
at the output of the optimal GQCM $M$ is a Gaussian state with correlation
matrix (CM) equal to%
\begin{equation}
V=\frac{1}{2}\left(
\begin{array}{cc}
(1+2\sigma ^{2})I & I \\
I & (1+1/2\sigma ^{2})I%
\end{array}%
\right) \text{ ,}  \label{CM}
\end{equation}%
where $I$ is the $2\times 2$ identity matrix. The CM of Eq.~(\ref{CM}) has
\emph{positive partial transpose} for every $\sigma ^{2}\geq 0$, and,
therefore, $\rho _{12}$ is always a separable state \cite{Simon}. This means
that Eve cannot exploit strategies based on the entanglement between her
clones and the ones of Alice and Bob. In the particular case of symmetric
cloning ($\sigma ^{2}=1/2$), we can write the useful decomposition%
\begin{gather}
\rho _{12}=\int d^{2}\mu \text{ }\Omega _{1/2}(\mu )\times  \notag \\
|\beta +\mu \rangle _{1}\langle \beta +\mu |\otimes |\beta +\mu \rangle
_{2}\langle \beta +\mu |\text{~.}  \label{decomposition}
\end{gather}%
Then, let us consider the case where the first cloner $M$ is optimal and
symmetric ($\sigma _{1}^{2}=\sigma _{2}^{2}=1/2$), while the second cloner $%
M^{\prime }$ is optimal but asymmetric, with $\sigma _{1^{\prime
}}^{2}\equiv \omega ^{2}$ and $\sigma _{2^{\prime }}^{2}=1/4\omega ^{2}$. In
this case, at the output modes $+$ and $-$ of the BS, we have the bipartite
state%
\begin{equation}
\rho _{+-}=\int d^{2}\mu \text{ }\Omega _{1/2}(\mu )\text{ }\chi (\mu )
\label{outputBS}
\end{equation}%
where%
\begin{align}
\chi (\mu )& \equiv \int d^{2}\lambda \text{ }\Omega _{1/4\omega
^{2}}(\lambda )\times  \notag \\
& |\theta _{+}+\lambda r\rangle _{+}\langle \theta _{+}+\lambda r|\otimes
|\theta _{-}+\lambda t\rangle _{-}\langle \theta _{-}+\lambda t|\text{ ,}
\label{Chi}
\end{align}%
and
\begin{equation}
\theta _{+}\equiv (\mu +\beta )(t+r)+\alpha r,\text{\ }\theta _{-}\equiv
(\mu +\beta )(t-r)+\alpha t\text{ .}  \label{theta}
\end{equation}%
If we now take a balanced BS (i.e., $t=r=1/\sqrt{2}$) we have $\theta
_{-}\equiv \alpha /\sqrt{2}$ and, therefore, the output port $-$ does no
longer contain the reference $\beta $. Here, the action of the BS\ is very
similar to the sum(mod2) performed over a binary key ($k$) and the
corresponding encrypted message ($k\oplus m$), operation that reveals the
message in the classical case ($k\oplus m\oplus k=m$). On the other hand,
the other port $+$ still contains a mixing between $\alpha $ and $\beta $
and, therefore, does not provide further information about the signal.
Tracing out this port, we have%
\begin{equation}
\rho _{-}=\int d^{2}\lambda \text{ }\Omega _{1/4\omega ^{2}}(\lambda )\text{
}|(\alpha +\lambda )/\sqrt{2}\rangle _{-}\langle (\alpha +\lambda )/\sqrt{2}|%
\text{ .}  \label{rho_m}
\end{equation}%
Heterodyning such a state, Eve can estimate the value of $\alpha $ up to a
Gaussian noise with variance
\begin{equation}
\sigma _{E}^{2}=2+(4\omega ^{2})^{-1}~,
\end{equation}%
for each quadrature. For Bob, instead, we have a total noise
\begin{equation}
\sigma _{B}^{2}=1+\sigma _{ch}^{2}~,
\end{equation}%
equal to the sum of the heterodyne noise ($1$) and the total channel noise
\begin{equation}
\sigma _{ch}^{2}=1/2+\omega ^{2}~.
\end{equation}%
According to Shannon, Bob ($B$) and Eve ($E$) will share with Alice ($A$) a
mutual information equal to $I_{AX}=\log (1+\gamma _{AX})$ with $\gamma
_{AX}\equiv \Sigma ^{2}/\sigma _{X}^{2}$ for $X=B,E$. Since \cite{Csiszar}
\begin{equation}
I_{AB}\geq I_{AE}\Longleftrightarrow \gamma _{AB}\geq \gamma
_{AE}\Longleftrightarrow \sigma _{B}^{2}\leq \sigma _{E}^{2}~,
\end{equation}%
we can easily compute a security threshold for this kind of attack, which is
equal to
\begin{equation}
\tilde{\sigma}_{ch}^{2}=(3+\sqrt{5})/4\simeq 1.3~.
\end{equation}%
Such a threshold must be compared with the security threshold ($0.5$) which
characterizes one-way coherent-state protocols \cite{Grangio,Christ} against
individual GQCM attacks.

\section{Conclusion}

In this paper we have considered one of the two-way protocols introduced in
\cite{Nat}. Then, we have explicitly studied its security in the presence of
particular kind of individual attacks which are based on combinations of
one-mode Gaussian cloners. Our analysis indicates that the superadditive
behavior of the security threshold should also hold against this kind of
Gaussian attacks. However, our analysis is far to be complete since we have
considered only particular combinations of cloners and we have also excluded
the possibility of a two-mode cloner (acting coherently on both the paths of
the quantum communication). Furthermore, the analysis covers the case of
direct reconciliation only. Despite these restrictions, the present work
represents the first step in the security analysis of two-way protocols
against more exotic kind of Gaussian interactions.

\section{Acknowledgements}

The research of S. Pirandola was supported by a Marie Curie Fellowship of
the European Community. S. Lloyd was supported by the W.M. Keck center for
extreme quantum information theory (xQIT).


\begin{thebibliography}{99}
\bibitem{Nat} S. Pirandola, S. Mancini, S. Lloyd, and S. L. Braunstein,
\textquotedblleft Continuous variable quantum cryptography using two-way
quantum communication,\textquotedblright\ Nature Physics \textbf{4}, 726
(2008).

\bibitem{SDiego} S. Pirandola, S. Mancini, S. Lloyd, and S. L. Braunstein,
\textquotedblleft Security of two-way quantum cryptography against
asymmetric attacks,\textquotedblright\ Proc. SPIE, Vol. 7092, 709215 (2008).
See also arXiv:0807.1937.

\bibitem{Ralph1} T. C. Ralph, \textquotedblleft Continuous variable quantum
cryptography,\textquotedblright\ Phys. Rev. A \textbf{61}, 010303(R) (2000);
T. C. Ralph, \textquotedblleft Security of continuous-variable quantum
cryptography,\textquotedblright\ Phys. Rev. A \textbf{62}, 062306 (2000);\
M. D. Reid, \textquotedblleft Quantum cryptography with a predetermined key
using continuous-variable Einstein-Podolsky-Rosen
correlations,\textquotedblright\ Phys. Rev. A \textbf{62}, 062308 (2000).

\bibitem{Preskill} D. Gottesman, and J. Preskill, \textquotedblleft Secure
quantum key distribution using squeezed states,\textquotedblright\ Phys.
Rev. A \textbf{63}, 022309 (2001); S. Iblisdir, G. Van Assche, and N. J.
Cerf, \textquotedblleft Security of quantum key distribution with coherent
states and homodyne detection,\textquotedblright\ Phys. Rev. Lett\textit{.}
\textbf{93}, 170502 (2004).

\bibitem{Grangio} F. Grosshans, G. Van Assche, J. Wenger, R. Brouri, N. J.
Cerf, and P. Grangier, \textquotedblleft Quantum key distribution using
Gaussian-modulated coherent states,\textquotedblright\ Nature \textbf{421},
238 (2003); F. Grosshans, and Ph. Grangier, \textquotedblleft Continuous
variable quantum cryptography using coherent states,\textquotedblright\
Phys. Rev. Lett. \textbf{88}, 057902 (2002).

\bibitem{Christ} C. Weedbrook \textit{et al.}, \textquotedblleft Quantum
cryptography without switching,\textquotedblright\ Phys. Rev. Lett. \textbf{%
93}, 170504 (2004); A. M. Lance \textit{et al.},\textit{\ }\textquotedblleft
No-switching quantum key distribution using broadband modulated coherent
light,\textquotedblright\ Phys. Rev. Lett. \textbf{95}, 180503 (2005).

\bibitem{Note} More properly, this channel is called \textit{additive
Gaussian noise channel}. For the general theory of these channels see, e.g.,
T. M. Cover and J. A. Thomas, \textquotedblleft Elements of Information
Theory\textquotedblright\ (Wiley, 2006).

\bibitem{Shannon} C. E. Shannon, \textquotedblleft A Mathematical Theory of
Communication,\textquotedblright\ Bell Syst. Tech. J. \textbf{27}, 623
(1948).

\bibitem{GQCM} N. J. Cerf, A. Ipe, and X. Rottenberg, \textquotedblleft
Cloning of continuous quantum variables,\textquotedblright\ Phys. Rev. Lett.
\textbf{85}, 1754 (2000).

\bibitem{Arthurs} E. Arthurs and J. L. Kelly, \textquotedblleft On the
simultaneous measurement of a pair of conjugate
observables,\textquotedblright\ Bell Syst. Tech. J. \textbf{44}, 725 (1965).

\bibitem{Shapiro} H.P. Yuen and J.H. Shapiro, \textquotedblleft Optical
communication with two-photon coherent states - part III: Quantum
measurements realizable with photoemessive detectors,\textquotedblright\
IEEE Trans. Inf. Theory \textbf{IT-26}, 78-92 (1980).

\bibitem{Simon} R. Simon, \textquotedblleft Peres--Horodecki separability
criterion for continuous variable systems,\textquotedblright\ Phys. Rev.
Lett. \textbf{84}, 2726 (2000).

\bibitem{Csiszar} I. Csisz\'{a}r and J. K\"{o}rner, \textquotedblleft
Broadcast channels with confidential messages,\textquotedblright\ IEEE
Trans. Inf. Theory \textbf{IT-24}, 339 (1978).
\end{thebibliography}
\end{document}